\documentclass[%
reprint,
 amsmath,amssymb,
 prl,
]{revtex4-1}

\usepackage{graphicx}
\usepackage{dcolumn}
\usepackage{bm}

\usepackage{comment}
\newcommand{\T}{\rule{0pt}{2.6ex}} 
\newcommand{\B}{\rule[-1.2ex]{0pt}{0pt}}

\begin{document}

\preprint{KCL-PH-TH-031}

\title{Reducing the Solar Neutrino Background in Dark Matter Searches Using Polarised Helium-3}

\author{Tarso Franarin}
 \email{tarso.franarin@kcl.ac.uk}
\author{Malcolm Fairbairn}
 \email{malcolm.fairbairn@kcl.ac.uk}
\affiliation{Physics, King's College London, Strand, London WC2R 2LS}

\begin{abstract}
Future dark matter detectors plan to have sensitivities such that solar neutrinos will start to become a problematic background. In this work we show that a polarised helium-3 detector would in principle be able to eliminate 98\% of these events when the orientation of the polarisation axis is antiparallel to the direction of the Sun.  We comment on the possible improvement in sensitivity of dark matter direct detection experiments due to this effect and the feasibility of building such a detector.
\end{abstract}

\pacs{Valid PACS appear here}
\maketitle


\section{Introduction}
A great deal of astronomical evidence suggests that approximately 25\% of the energy density of the Universe today is composed of cold, non-baryonic ``dark matter'' (DM) \cite{planck,particledm}. Amongst the plethora of dark matter candidates, thermal relic particles with self-annihilation cross sections set by the electroweak scale are thought to be well motivated as they can lead naturally to the observed abundance and because the hierarchy problem suggests that we expect new physics to show up at the energy scale. Direct detection experiments aim to detect such particles via the nuclear recoils caused by their elastic scattering off nuclei \cite{dd,guo}.

In the near future, the target masses of DM detectors will be increased to the ton-scale \cite{ton}, and neutrino-nucleon scattering from solar neutrinos will become detectable \cite{billard}.  This can be viewed as a positive development, potentially shedding light on new aspects of solar physics and neutrino physics beyond the standard model \cite{strigari,davissn,cerdeno}.  However it is a hindrance to the potential detection of light dark matter because these solar neutrinos act as a background which produce signals very similar to dark matter candidates. In recent years, several strategies have been suggested to distinguish between neutrino and dark matter signals. These involve directional detectors \cite{philipp,ohare}, annual modulation of both the dark matter and solar neutrinos signals \cite{davis}, data combination from detectors composed of different target materials \cite{ruppin}, and detectors with improved energy resolution \cite{dent}.

In this work we present another strategy, that of a polarised helium-3 target where the cross section for neutrino scattering is a strong function of the angle between the neutrino momentum and the axis of polarisation. Assuming the detector is not close to a reactor, most low energy background neutrinos come from the Sun. We analyse the dependence of the neutrino signal distribution to the orientation of the incoming neutrinos in a polarised detector and use it to reduce this background.

\begin{figure}[b]
    \centering
    \includegraphics[width=0.46\textwidth]{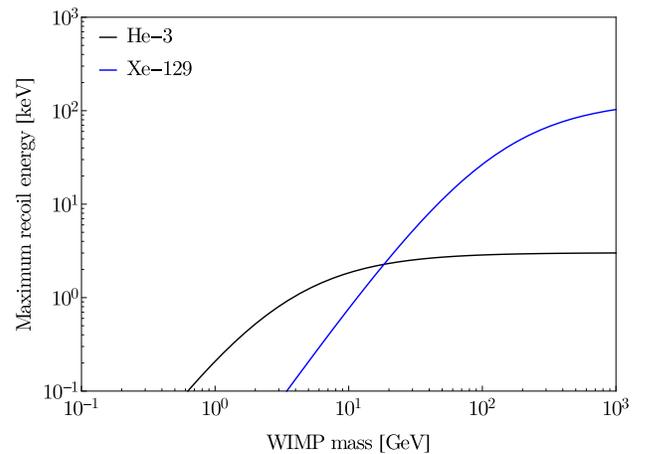}
    \caption{Maximum recoil energy of the target nucleus versus the incident
WIMP mass for helium-3 and xenon-129. For helium-3, the recoil energy range for all the possible dark matter events is higher-bounded.}
    \label{fig:ermax}
\end{figure}

\section{Helium-3}

The use of helium-3 for dark matter detection is very appealing for many reasons \cite{moulin1,moulin2}: being a spin 1/2 nucleus, it is sensitive to the axial interactions with WIMPs; for massive WIMPs the maximum recoil energy depends very weakly on its mass and, therefore, is higher-bounded (see figure \ref{fig:ermax}); it has a low Compton cross section to $\gamma$-rays reducing by several orders of magnitude the natural radioactive background; it has no intrinsic X-rays; the capture process $n+ ^3\text{He}\rightarrow p+ ^3\text{H}+\text{764 keV}$ gives a clear signal for neutron rejection; it can be polarised and used as a target \cite{jlab}. 

There are two leading methods to polarise a significant amount of nuclei in gaseous helium-3: spin-exchange optical pumping (SEOP) \cite{walker,bouchiat} and metastability exchange optical pumping (MEOP) \cite{batz,colegrove}. The latter method is not well suited for our purposes because it needs very low pressure ($\approx$ 1 mbar), implying a huge volume for the detector. In the first method, which can operate at typical pressures of 1 - 10 bar, alkali metal vapour is polarised by optical pumping and then transfers its electronic polarisation to the helium-3 nuclei via spin-exchange collisions. Originally, a pure rubidium vapor was used in this process. However, since potassium is much more efficient than rubidium at transferring its polarisation to helium-3 nuclei, hybrid mixtures of these elements are used, leading to degrees of polarisation up to 70\% \cite{singh}. 

We note that the use of any such methods might risk contamination of the final helium-3 gas with potassium or rubidium isotopes. The use of other alkalis have been proposed, including non-radioactive elements such as sodium \cite{babcock}.

\section{Neutrino Background}

A standard model neutrino is thought to scatter simultaneously off all nucleons in a nucleus in phase when the wavelength of the momentum transfer is larger than the radius of the target nucleus. This coherent neutrino-nucleus scattering leads to a cross section enhanced by a factor of $[N-(1-4\text{sin}^2\theta_W)Z]^2$, where $N$ and $Z$ are the number of target neutrons and protons respectively, and $\theta_W$ is the weak mixing angle \cite{freedman}. This process has never been observed in standard neutrino detectors due to the tiny cross section and very low nuclear recoil energies.  The coherent nature of the interaction depends upon the vectorial $V$ part of the $V-A$ standard model coupling. The axial $A$ part also leads to a scattering upon any net spin of the nucleus, but this will not be coherently enhanced since in general there will be only one unpaired nucleon.  For a target like xenon for example, the axial coupling will give rise to a much smaller spin-dependent cross-section than the coherent spin-independent cross section arising from the vector current.

The maximum recoil kinetic energy in coherent neutrino-nucleus scattering is
\begin{equation}
E_{r,\text{max}}=\frac{2E_\nu^2}{m_N+2E_\nu},
\end{equation}
where $E_\nu$ is the incident neutrino energy, and $m_N$ is the mass of the target nucleus. The three-momentum exchange is related to the recoil energy by $q\approx\sqrt{2m_NE_r}$. For neutrino energies below 20 MeV and nuclear targets from ${}^3$He to ${}^{132}$Xe, the maximum recoil energy ranges between 280 keV and 6 keV, meaning that the maximum possible $q$ is quite small, $<$ 1 $\text{fm}^{-1}$. Typical nuclear radii, $R$, are 3-5 fm, then the product $qR<1$. In this regime, the neutrino scatters coherently off the nucleus.

The most important contribution to the neutrino flux in the lowest energy range capable of giving a detectable nuclear recoil (around a keV) comes from the various nuclear fusion and decay processes occurring in the Sun. For this work, we consider only the fluxes of solar neutrinos. Atmospheric neutrinos and neutrinos from the diffuse supernova background are produced at higher energies and with much lower rates, only being detectable in future multi-ton detectors \cite{billard}.

\begin{figure}
    \centering
    \includegraphics[width=0.48\textwidth]{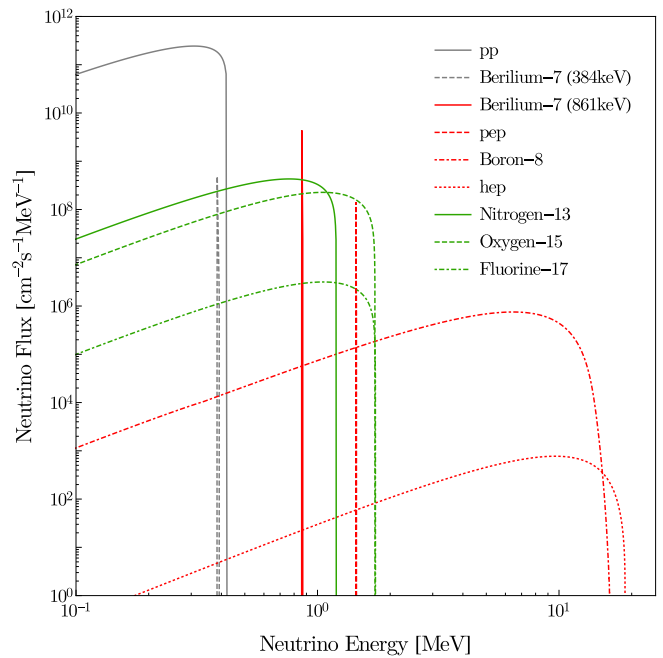}
    \caption{Solar neutrinos fluxes. The grey colored fluxes will not give events above thresholds considered in this paper. Solar neutrinos are produced in the proton-proton chain reaction (red) and CNO cycle (green).}
    \label{fig:fluxes}
\end{figure}

In figure \ref{fig:fluxes} we show the solar neutrino fluxes. Events induced by neutrinos from the grey coloured fluxes will not give events above the threshold considered in this paper. Solar neutrinos produced in the proton-proton chain are in red, while those produced in the CNO cycle are in green. 

\section{Neutrino Background Distribution}

The differential cross section for a neutrino scattering off a polarised nucleon is

\begin{equation}
\begin{split}
\frac{d\sigma}{d\Omega}=&\frac{G_F^2E_\nu^{\prime 2}}{16\pi^2E_\nu^2m_N^2}\bigg\{c_V^2\Big\{(pp_N)(p^{\prime}p_N^{\prime})+(pp_N^{\prime})(p^{\prime}p_N)\\&+m_N\Big[(pS)[(p^{\prime}p_N)-(p^{\prime}p_N^{\prime})]-(p^{\prime}S)[(pp_N)-(pp_N^{\prime})]\Big]\\&-m_N^2(pp^{\prime})\Big\}+c_A^2\Big\{(pp_N)(p^{\prime}p_N^{\prime})+(pp_N^{\prime})(p^{\prime}p_N)\\&-m_N\Big[(pS)[(p^{\prime}p_N)+(p^{\prime}p_N^{\prime})]-(p^{\prime}S)[(pp_N)+(pp_N^{\prime})]\Big]\\&+m_N^2(pp^{\prime})\Big\}+2c_Vc_A\Big\{(pp_N)(p^{\prime}p_N^{\prime})-(pp_N^{\prime})(p^{\prime}p_N)\\&-m_N\Big[(pS)(p^{\prime}p_N^{\prime})+(p^{\prime}S)(pp_N^{\prime})\Big]\Big\}\bigg\}.
\end{split}
\end{equation}

\noindent Here $G_F$ is the Fermi constant, $E_\nu$ is the incoming neutrino energy, $E_\nu^\prime$ is the outgoing neutrino energy, and $c_V$ and $c_A$ are the effective couplings. The neutrino and nucleus initial (final) four-momenta are respectively $p$ ($p^{\prime}$)  and $p_N$ ($p_N^{\prime}$), and $S$ is the nuclear spin four-vector.  Because the recoil energies are much less than the nuclear masses and neutrino energies, this expression reduces to
\begin{equation}
\begin{split}
\frac{d\sigma}{d\Omega}=&\frac{G_F^2E_\nu^2}{16\pi^2}\{c_V^2+3c_A^2+(c_V^2-c_A^2)\text{cos}\psi\\&-2c_A[(c_V-c_A)\hat{v}.\hat{s}+(c_V+c_A)\hat{v}^\prime.\hat{s}]\},
\end{split}
\label{eq:dsdo}
\end{equation}
where $\psi$ is the scattering angle in the lab frame of the outgoing neutrino with respect to the incoming neutrino direction, $\hat{v}$ and $\hat{v}^\prime$ are respectively the directions of the incoming and outgoing neutrinos and $\hat{s}$ is the direction of the nuclear spin.
\begin{figure*}
\begin{minipage}[t]{0.48\linewidth}
\includegraphics[width=.9\linewidth]{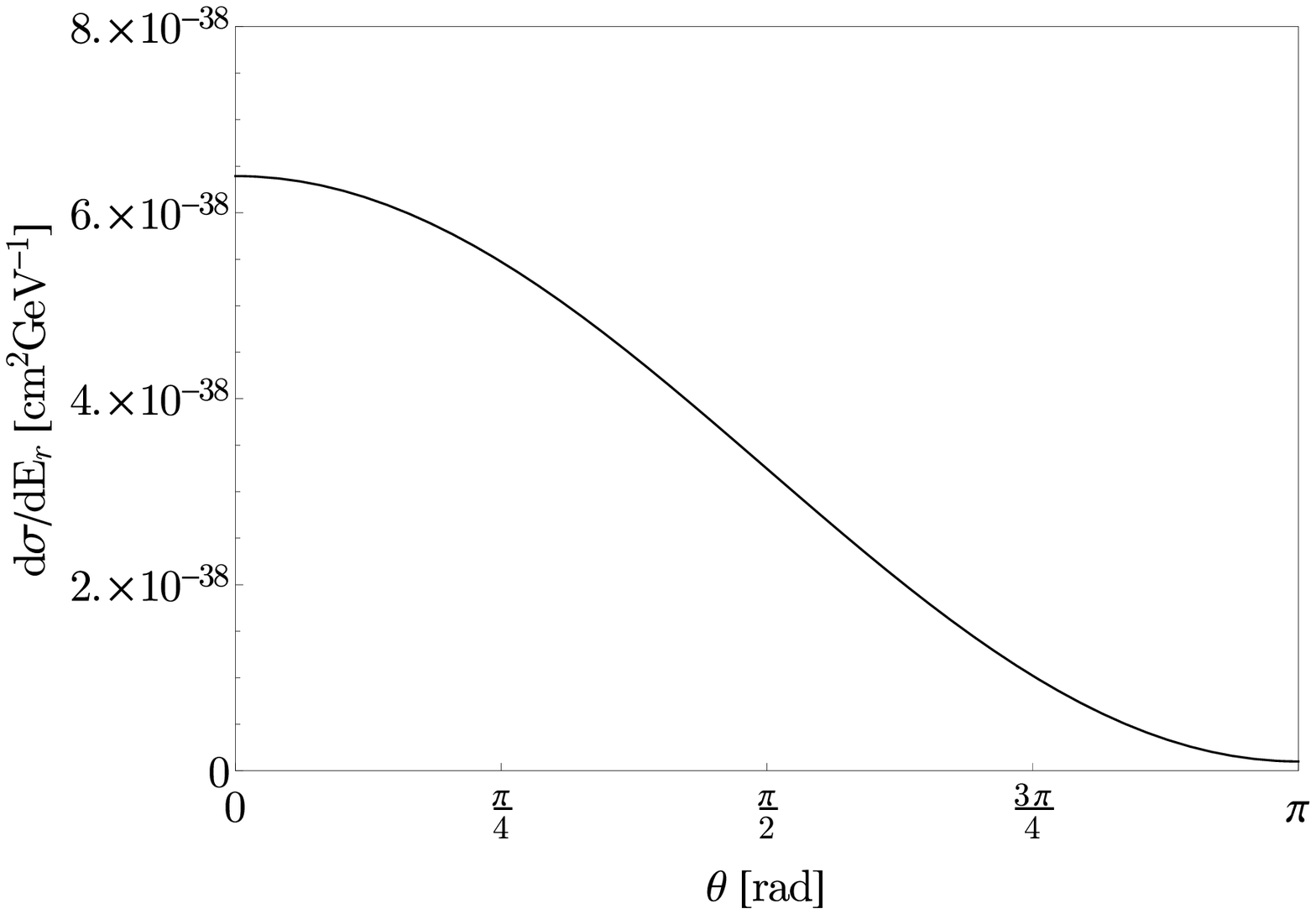}
\end{minipage}\hfill
\begin{minipage}[t]{0.48\linewidth}
\includegraphics[width=.9\linewidth]{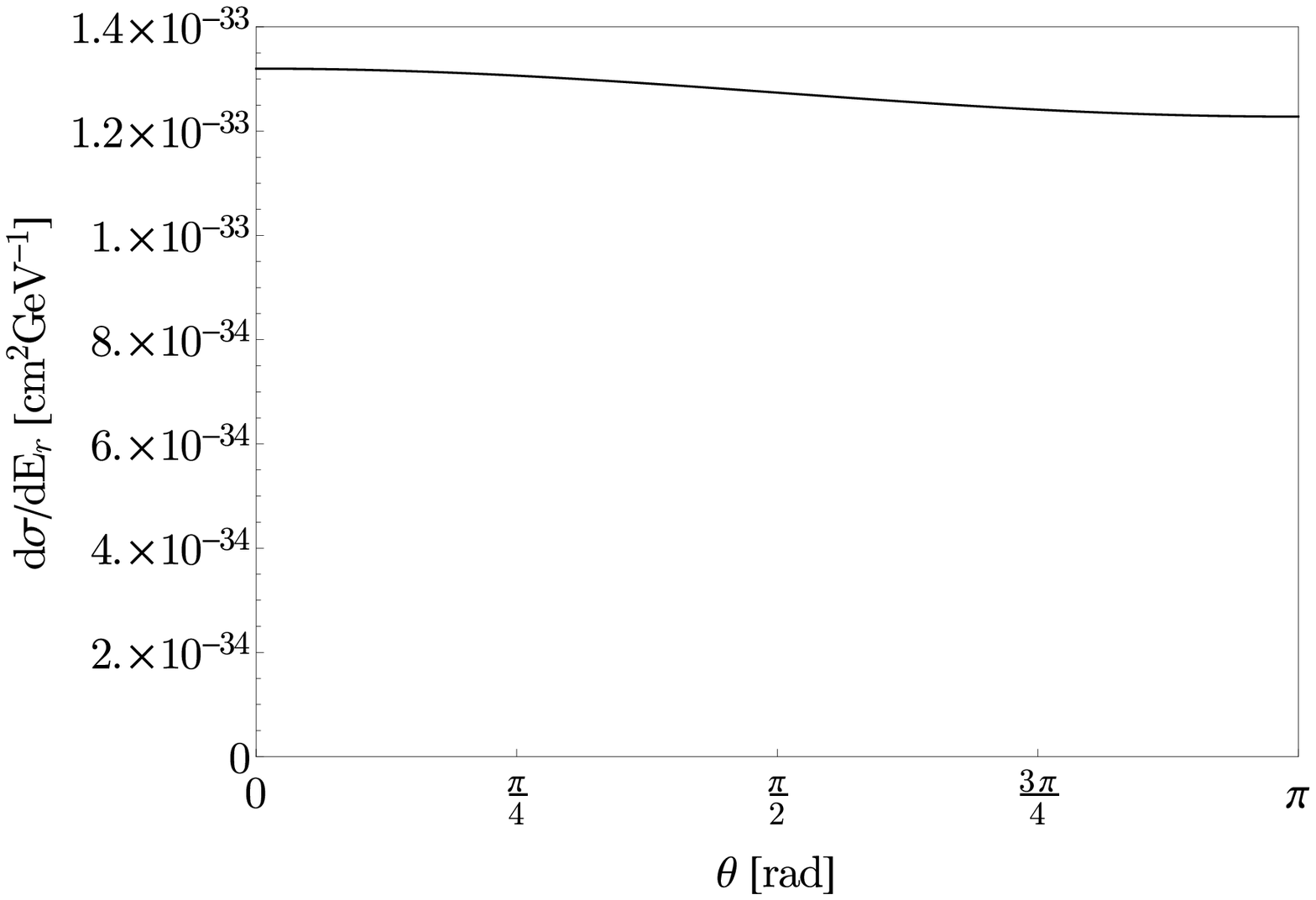}
\end{minipage}
\caption{Neutrino-nucleus differential cross section for helium-3 (left) and xenon-129 (right) for $E_\nu$=6.4 MeV and $E_r=E_{r,\text{max}}/2$ as a function of $\theta$, the angle between the incoming neutrino and the nuclear spin directions.}
\label{fig:cross}
\end{figure*} 
For low-energy interactions, atomic nuclei behave, in a good approximation, as a collection of independent nucleons, which yields
\begin{equation}
c_V^\text{nucleus}=Zc_V^p+Nc_V^n
\end{equation}
and
\begin{equation}
c_A^\text{nucleus}=c_A^\text{unpaired nucleon}
\label{eq:ca}
\end{equation}
for a DM-nucleus coupling, where $c_V^{p,n}$ and $c_A^{p,n}$ are respectively the effective vector and axial neutrino couplings to protons and neutrons. We take the values shown in Table \ref{tab:couplings}. In equation \ref{eq:ca} it is assumed that the nuclear spin is solely due to the spin of the single unpaired nucleon.

\begin{table}[h!]
\centering
\begin{tabular}{c|c|c}
& $c_V$ & $c_A$ \B\\ \hline
Proton & $1-4\text{sin}^2\theta_W$ & 1.26 \T\B\\ \hline
Neutron & -1  & -1.26 \T\\
\end{tabular}
\caption{Effective neutral-current couplings.}
\label{tab:couplings}
\end{table}

From scattering kinematics, it can be found that
\begin{equation}
\text{cos}\psi=1-\frac{E_r}{m_N}\left(\frac{E_\nu+m_N}{E_\nu}\right)^2.
\label{eq:psi}
\end{equation}
Then it follows from that
\begin{equation}
\frac{d\sigma}{dE_r}=\int_0^{2\pi}\frac{d\sigma}{d\Omega}\frac{d(\text{cos}\psi)}{dE_r}d\phi=\frac{2\pi}{m_N}\left(\frac{E_\nu+m_N}{E_\nu}\right)^2\frac{d\sigma}{d\Omega}.
\end{equation}

Figure \ref{fig:cross} shows $\frac{d\sigma}{dE_r}$ for a neutrino scattering off a helium-3 and a xenon-129 nucleus, the most abundant spin-1/2 xenon isotope. In the helium-3 case the coherent spin-independent contribution doesn't overshadow the spin-dependent one.
 We can quantify the magnitude of this effect considering the relative amplitude

\begin{equation}
\alpha=\frac{1}{2}\left|\frac{\frac{d\sigma}{dE_r}(0)-\frac{d\sigma}{dE_r}(\pi)}{\frac{d\sigma}{dE_r}(\pi/2)}\right|.
\end{equation}
Its value ranges from 0 (no angular dependence) to 1 (maximum angular dependence). Table \ref{tab:amp} below shows the values of $\alpha$ for different elements taking $E_\nu$=6.4 MeV and $E_r=E_{r,\text{max}}/2$.

\begin{table}[h!]
\centering
\begin{tabular}{c|c}
& $\alpha$ \\ \hline 
\T
$^3$He & 0.97 \T\B \\ \hline 
$^{13}$C & 0.41\T\B \\  \hline 
$^{15}$N & 0.36 \T\B\\  \hline 
$^{19}$F & 0.22\T\B \\  \hline 
$^{129}$Xe & 0.04\T \\
\end{tabular}
\caption{Relative amplitude of the differential cross section for different elements.}
\label{tab:amp}
\end{table}

The number of neutrino events is calculated as an integral over the differential rate and the energy dependent detection efficiency $\epsilon(E_r)$:
\begin{equation}
N_\nu=\int_{E_{\text{thr}}}^{E_{\text{max}}}\epsilon(E_r)\frac{dR_\nu}{dE_r}dE_r.
\end{equation}
The differential rate is
\begin{equation}
\frac{dR_\nu}{dE_r}=n_T\; T\int_{E_\nu^\text{min}}^{E_\nu^\text{max}}\frac{dN_\nu}{dE_\nu}\frac{d\sigma(E_\nu,E_r,\theta)}{dE_r}dE_\nu,
\end{equation}
with $n_T$ the number of target nuclei in the detector, $T$ is the exposure time, $\frac{dN}{dE_\nu}$ the neutrino flux, and $\frac{d\sigma}{dE_r}$ the differential cross section, where $\theta$ is the angle between $\hat{v}$ and $\hat{s}$. Rigorously, the solar neutrino flux depends on the annual variation in the Earth-Sun distance. The flux of solar neutrinos vary by around 3\% over a year and the shape of the spectrum is not affected. As a first approximation, we take this flux to be time-independent.

The total number of solar neutrinos per 100 kg-year of helium-3 assuming ideal energy efficiency is shown in Table \ref{tab:ne}. When the detector is fixed at $\theta=\pi$, this number is very low. We can then consider a polarised helium-3 dark matter detector that keeps the angle $\theta$ fixed at $\pi$ to minimise the number of solar neutrinos events. 

For a number of spin 1/2 nuclei, the level of polarisation $P$ is defined as
\begin{equation}
P=\frac{N^\uparrow-N^\downarrow}{N^\uparrow+N^\downarrow},
\end{equation}
where $N^\uparrow$ and $N^\downarrow$ are respectively the number of nuclear spin states +1/2 and -1/2. 

In a 100 kg unpolarised detector with a threshold energy of 0.2 keV, around 43 solar neutrinos are expected to be detected every year. The same detector with a 70\% degree of polarisation and fixed optimal angle would detect roughly 13 solar neutrinos each year. In the ideal case where this detector is completely polarised, 1 solar neutrino would be expected each year.
 
\begin{table}[h!]
\centering
\begin{tabular}{c|c|c|c|c}
$E_\text{thr}$ (keV) &2 & 1 & 0.5 &0.2 \B\\ \hline
$N_{\text{E}}(0)$  & 3.23& 7.67 & 23.26& 85.45 \T\B \\ \hline
$N_{\text{E}}(\pi)$  & 0.07& 0.10  &0.19& 0.85 \T\B\\ \hline
$N_{\text{E,unpolarised}}$  & 1.65& 3.89 &11.73& 43.15 \T\\
\end{tabular}
\caption{Number of solar neutrino events per 100 kg-year of helium-3.}
\label{tab:ne}
\end{table}

\section{Dark Matter}

For dark matter particles the polarisation modulated amplitude is small because it is velocity suppressed \cite{kamionkowski}, so the detector's orientation is largely irrelevant for those events, even though there is a preferred arrival direction for the fastest dark matter particles due to the motion of the solar system through the Galaxy.

Direct dark matter detection results put tight constraints on a $Z^\prime$ with vector couplings to light quarks due to the coherent enhancement in such models. Conversely, models where dark matter scattering on nuclei is spin-dependent are very weakly constrained \cite{kg,axialdm}. For such a class of models, helium-3 detectors would be more efficient than xenon-based detectors since 1 kg of helium-3 has the same number of unpaired neutrons as 90 kg of xenon.

\begin{figure}[b]
    \centering
    \includegraphics[width=0.49\textwidth]{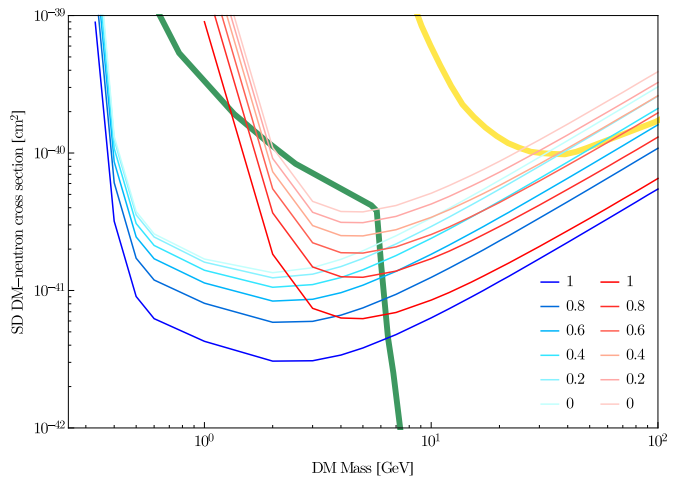}
    \caption{Spin-dependent exclusion curve for an exposure of 100 kg-year, threshold energies of 0.2 keV (blue) and 2 keV (red), and different levels of polarisations.  The neutrino floor for xenon (green) has been rescaled by the appropriate $2A^2$ factor, where the factor of two accounts for the fact that roughly half of xenon nuclei have an unpaired neutron. In yellow, the 90\% C.L. limit on SD DM-neutron cross section from LUX \cite{lux}.}
    \label{fig:exclusion}
\end{figure}

Figure \ref{fig:exclusion} shows the limits on the spin dependent cross section that we expect to be able to obtain with an exposure of 100 kg-year for a helium-3 detector for thresholds of 0.2 keV (blue) and 2 keV (red).  One can expect gamma ray backgrounds with very good shielding of about 1 event keV$^{-1}$kg$^{-1}$day$^{-1}$. To go below this one needs advanced strategies for background rejection, for example self-shielding or analysis of charge to light ratio in xenon detectors.  The limits come from performing Monte Carlo simulations of distinguishing signal (DM+neutrino) from background (neutrino only) over bins of 0.2 keV width with an upper energy threshold limit of 20 keV.  The addition of polarisation increases the sensitivity by nearly an order of magnitude.  In the figure, we don't assume any particular coherent enhancement between dark matter and the nuclei, so the limits represent the spin-dependent constraint, although any coherence effects would be minor for helium-3.  

This figure looks quite different to normal dark matter sensitivity plots where the neutrino floor due to solar neutrinos for low mass dark matter is present.  The reason for this is that for such a light target, the recoil energies expected from dark matter are very small indeed, while the recoil energies from the solar neutrinos are much larger, resulting in neutrinos being a background for much heavier dark matter. At the masses we plot here, the neutrino background includes contibutions from many different reactions in the Sun.  The result is an overall diminishment of sensitivity, rather than one which is only significant at low dark matter masses.

\section{Conclusions}

In this work we looked at the neutrino floor problem for future direct dark matter searches, when neutrino backgrounds from coherent neutrino-nucleus elastic scattering are important. These neutrinos interact either via the vector or axial current, the former of which depends roughly on $N^2$, where N is the number of neutrons, and the latter of which depends on the net spin of the nucleus. Hence for lighter elements such as helium-3 the two currents, labeled spin-independent and spin-dependent respectively, are of similar size. 

We investigated how a polarised detector can in theory help reduce this background. We considered a detector based on helium-3 so that the spin-dependent contribution is not suppressed relative to the coherent cross section and the effect of polarisation is maximised.

We showed that the neutrino-nucleus cross section has a strong dependence on the angle between the incoming neutrino for helium-3 and the nuclear spin. This property can be used in the construction of a polarised detector, which is kept at an angle that minimises the number solar neutrinos detected. Such a detector operating at full polarisation would be able to rule out 98\% of solar neutrino events. For dark matter this effect is negligible since it's velocity suppressed and these particles are non-relativistic.

We showed that such discrimination could reduce the neutrino background from the Sun for light dark matter by almost an order of magnitude and performed estimates of the potential performance of such a detector with and without polarisation for different energy thresholds.

The cost of helium-3 is very high relative to other materials which are used for dark matter detection, and we are aware of the difficulties that would be encountered in setting up such an experiment in practice.  The purpose of this note was to point out the nice features that such a detector would have if we were able to polarise the nuclei, especially with regards to their interactions with neutrinos.  We note that this nice feature of helium-3 comes hand in hand with higher recoil energies for neutrino events, meaning there are more neutrino events above a certain threshold.  Detailed studies to optimise this trade off and perhaps to look at other target nuclei would be natural extensions of this work.
Finally, a drop in price of helium-3 would be helpful in making the realisation of such a device feasible.

\section*{Acknowledgment}
We are extremely grateful for comments from Chamkaur Ghag, Mark Kamionkowski, Dan McKinsey, and Jim Wild. TF thanks support from CNPq SwB grant. MF acknowledge support from the STFC and funding from the European Research Council under the European Union's Horizon 2020 program (ERC Grant Agreement no.648680 DARKHORIZONS).


\end{document}